\documentclass[aps,prl,twocolumn, superscriptaddress, groupedaddress]{revtex4}  
\usepackage{graphicx}  
\usepackage{dcolumn}   
\usepackage{bm}        
\usepackage{amssymb, amsmath}
\usepackage{wasysym}

\def\beq{\begin{equation}}
\def\eeq{\end{equation}}
\usepackage[usenames,dvipsnames,svgnames]{xcolor}
\usepackage{hyperref}
\hypersetup{colorlinks, citecolor=black, linkcolor=black, urlcolor=black}

\usepackage{textcomp}
\usepackage{amsmath}
\usepackage{amsfonts}
\usepackage{amssymb}
\usepackage{bbm}                
\usepackage{graphicx, rotating}
\usepackage{epstopdf}
\usepackage{epsfig}
\usepackage{latexsym}
\usepackage{graphicx}
\usepackage{bbold}
\usepackage{color}
\usepackage{amsmath,amssymb}

\newcommand{\bea}{\begin{eqnarray}}
\newcommand{\eea}{\end{eqnarray}}
\newcommand{\eq}[1]{Eq.~(\ref{#1})}

\begin{document}
\widetext
\begin{flushright}
	 IPPP/18/37
\end{flushright}

\title{The  Relaxion Measure Problem}
\author{Rick S. Gupta}
\affiliation{Institute of Particle Physics Phenomenology, Durham University, Durham, UK DH1 3LE}
\begin{abstract}
We examine the necessity of requiring that relaxion dynamics is dominated by classical slow roll and not quantum fluctuations. It has been recently proposed by Nelson and Prescod-Weinstein~\cite{Nelson}  that abandoning this requirement  can lead to a  unified solution of the hierarchy and strong CP problems in QCD relaxion models. In more general models this results in a higher value of the allowed cut-off. In this work we find, however, that relaxing this condition and can result in the universe being dominated in physical volume by regions arising from large quantum fluctuations of the relaxion.  These regions turn out to be problematic for the relaxion mechanism because either the relaxion does not stabilise at all or it stabilises  at vacua which cannot reproduce the observed properties of our universe. The size of these undesirable regions is moreover ambiguous because of the measure problem. For instance, we show that if one chooses to use the scale factor cut-off measure  such dangerous regions occupy a negligible volume and  these issues do not arise. 
\end{abstract}
\maketitle

The relaxion mechanism, proposed by  Graham, Kaplan and Rajendran (GKR)~\cite{gkr}, presents a third way of addressing the electroweak hierarchy problem that uses neither symmetries nor anthropics.  The Higgs boson mass in these models is scanned by a slowly rolling field during inflation.  The scanning stops when the Higgs mass is close to zero because of a feedback mechanism thus explaining the hierarchy between the electroweak scale and a much higher cut-off scale.  

A nice feature of relaxion models is that it can be unified with solutions to  other naturalness problems like the strong CP problem and the  the Standard Model flavour puzzle~\cite{Nelson,gkr,nbr,hir}. In the very first relaxion model proposed by GKR in Ref.~\cite{gkr}, the QCD axion is itself the relaxion. As the relaxion stops at an ${\cal O}(1)$ phase, however, this model gives an an ${\cal O}(1)$ value for the strong CP phase, $\theta_{\rm QCD}$, and is thus ruled out. An elegant improvement of this model has been recently  proposed by Nelson and Prescod-Weinstein (NP)~\cite{Nelson}. In this work the authors show that the original GKR model  can be compatible with the experimental constraints on the strong CP phase if the Hubble scale during inflation is   larger than the QCD scale. As we will soon describe in more detail, this leads to a suppression of the axion potential during inflation but as the Hubble scale becomes smaller at the end of inflation, the axion potential becomes larger, and the relaxion stabilises at a very small value of $\theta_{\rm QCD}$.   

Taking a large Hubble scale, however, means that one has to relax the condition that the  Hubble induced quantum fluctuations of the relaxion field are small enough so that its dynamics can be approximated to be classical.  As we discuss in the present work allowing such  large quantum fluctuations of the relaxion field can be problematic. This is despite the fact that regions where the relaxion spreads far from its classical (expectation) value can be shown to be exponentially small in volume during inflation. The subtleties arise after the inflaton stabilises when some of the above regions  expand exponentially because they  have a high energy density. As we will show, this exponential expansion can potentially compensate for the initial exponential suppression if the so called quantum vs classical (QvsC) requirement is not imposed.  To  know whether this can spoil the relaxion mechanism, one needs to compute ratio of the volume where the relaxion dynamics successfully explains a small weak scale (and in the case of the NP model a small $\theta_{\rm QCD}$) to the volume generated by these large fluctuations where this might not be achieved. There is no unambiguous way to carry out this computation, however, because both these volumes are generally infinite. The only way to regulate these infinities depends on how we choose time slices across causally disconnected regions of spacetime which is ultimately arbitrary. This is a statement of the the `Relaxion Measure Problem' that we will explain in more detail in what follows.

Our conclusions would be relevant not just for the NP model but also for more general relaxion models where relaxing the QvsC requirement leads to a larger value of the allowed cut-off.  More importantly it conceptually clarifies the need for the QvsC requirement in general relaxion models.  Many of the issues raised in this paper were already qualitatively  anticipated in Ref.~\cite{Nelson,gkr}. The purpose of this work is to examine these ideas in more quantitative detail and take them to their logical conclusion. In particular we confirm the suggestion of NP that these problems do not arise if one uses the scale factor cut off measure.

\section{Review of the relaxion mechanism}

Let us first present a very brief review of the  relaxion mechanism focussing especially on the  models of GKR  and NP where the relaxion is also the QCD axion. In relaxion models the value of $\mu^2$, the  mass squared term in the Higgs potential, changes during the course of inflation as it depends on the relaxion, $\phi$,
\bea
V(H,\phi)&=&\mu^2(\phi) H^\dagger H+\lambda (H^ \dagger H)^2+\Lambda_c-g M^2 \phi+...\nonumber \\
\mu^2(\phi)&=&M^2-g \phi + ...   \, ,
\label{linear}
\eea
which slowly rolls because of the potential due to the linear potential above \footnote{Here the ellipses refer to higher dimensional terms such as $g^2 \phi^2 , g^3 \phi^3/M^2, g^2 \phi^2 H^ \dagger H/M^2$ etc. These contribute at the same order as the linear term for $\phi\simeq \phi_c \simeq \frac{M^2}{g}$ and thus ignoring them does not affect the above analysis.}.  Here $g$ is a dimension-full coupling  and $M$ is the  scale where the Higgs quadratic divergence gets cut off. The field $\phi$ starts rolling from an initial  field value $\phi < {M^2}/{g}$, such that $\mu^2$ is positive and electroweak symmetry is unbroken. After crossing the point, $\phi_c= {M^2}/{g}$,  $\mu^2$ becomes negative and the Higgs gets a vacuum expectation value (VEV),  $v^2(\phi)={-\mu^2(\phi)}/{\lambda}$. This triggers the so called backreaction potential that leads to the barriers in Fig.~1,
  \begin{eqnarray}
   \Delta V_{br}(h,\phi)\simeq -\Lambda_{br}^4 \cos \left( \frac{\phi}{f} \right),
   \label{br}
  \end{eqnarray}
  where where $\Lambda_{br}^4=m^j v^{4-j}$ with $0\leq j \leq 4$. In the QCD relaxion models where the relaxion is also the QCD axion and has the coupling
\beq
 \frac{\phi}{f} G_{\mu \nu} \tilde{G}^{\mu\nu},
\eeq
the non-zero Higgs VEV turns on the leading term of the zero temperature axion potential, 
  \begin{eqnarray}
   \Lambda_{br}^4\simeq y_u v f_\pi^3,
   \label{brqcd}
  \end{eqnarray}
where $y_u$ is the Yukawa coupling of the up quark. As $\phi$ continues rolling, $|\mu^2(\phi)|$ becomes larger, resulting in a monotonically increasing Higgs VEV, thus  increasing the size of the barriers. Eventually the barriers become large enough and the relaxion stops rolling  at an arbitrary ${\cal O}(1)$ value of the phase $\phi_0/f$ where $\partial_\phi V(h,\phi)=0$,
  \begin{eqnarray}
g M^2=\frac{\Lambda_{br}^4}{f} \sin \left( \frac{\phi_0}{f} \right).
 \label{deriv}
  \end{eqnarray}   
The ${\cal O}(1)$  phase $\phi_0/f$  is precisely  $\theta_{\rm QCD}$ and thus this model is ruled out by experiments which require $\theta_{\rm QCD}<\theta_{\rm QCD}^{\rm ub}\sim10^{-10}$~\cite{Vicari,Afach}. We will soon discuss how the NP model attempts to resolve this issue.  If $g$ is small enough the cut-off can be raised much above the electroweak scale. Note that a small  $g$ 
 is radiatively stable as in the limit  $g\to 0$,  the discrete symmetry $\phi\rightarrow \phi+2\pi k f~(k\in \mathbb{Z}) \,$ is recovered.    As it was pointed out in Ref.~\cite{GuptaK}, however, the coupling $g$ is very problematic because  the Peccei Quinn (PQ)  axion is usually identified with the angular part of a complex scalar field having a periodicity $2 \pi f$, making the non-periodic terms proportional to $g$ impossible. The only known way of resolving this issue is to imagine that the PQ symmetry has a large non anomalous discrete subgroup $Z_N$ so that the periodicity that appears in the axion potential is smaller than the original periodicity, say $F$,  by a factor $N$, i.e. $f=F/N$~\cite{GuptaK} (see also Ref.~\cite{servant1}). However, as we must have $F$  larger than the typical distance, $M^2/g$, the relaxion travels, \eq{deriv}  implies that we need an extremely large $N\sim M^4/\Lambda_{br}^4$. Such large values of  $N$ can be succesfully realized  in the so called clockwork models with multiple axions~\cite{cky, ci, kapr}. In such a scenario the the Lagringian terms proportional to $g$ are also secretly periodic but with a much larger periodicity, 
\begin{eqnarray}
V(H,\phi)&=&\mu^2(\phi) H^\dagger H+\lambda (H^ \dagger H)^2+\kappa_2 M^{4}\cos\frac{\phi}{F}+\Lambda_c\nonumber \\
\mu^2(\phi)&=&-M^2+\kappa_1 M^2 \cos \frac{\phi}{F}
\label{cosine}
\end{eqnarray}
where we must have $\kappa_{1}\gtrsim 1$ to ensure $\mu^2$ changes sign during the relaxion slow roll  and $\kappa_2\sim 1$. We can recover \eq{linear} from \eq{cosine} by expanding about $\phi_0=\pi F/2$,  identifying,
\beq
g \sim M^2/F.
\label{interchange}
\eeq
and redefining $\Lambda_c$. As we will see, taking the rolling potential to be \eq{cosine} instead of \eq{linear}  will be crucial for another reason: a periodic potential  puts an automatic upper bound on the size of quantum fluctuations of  $\phi$.  Otherwise, \eq{linear} and \eq{cosine} are equivalent in the sense that they yield the same results parametrically. In particular, unless the above mentioned subtleties are relevant,  the equations in this paper can be written interchangeably (up to  ${\cal O}(1)$ factors) in  terms of either $g$ or $F$  using \eq{interchange}. 

We now list the conditions GKR imposed on the parameter space to ensure that the above picture is consistent cosmologically. First of all, we must assume that the relaxion energy density is a negligible contribution to the total energy density during inflation, i.e., 
\beq
M^4\lesssim H_I^2 M_{pl}^2
\label{cond1}
\eeq
 so that the Hubble scale, $H_I$, and other details related to inflation are independent of relaxion dynamics. Here $M_{Pl}$ is the reduced Planck mass.  In this work we will assume that the Hubble expansion rate $H_I$ is due to the vacuum energy of a single inflaton  $\sigma$. Furthermore as we would be interested only in the large quantum fluctuations of the relaxion and not the inflaton, we will assume as in the original   GKR paper, that the  dynamics of the inflaton is dominated by classical slow roll. Now we come to the all important Quantum vs Classical (QvsC)  condition. So far the description of the dynamics of $\phi$ has been completely classical.  If quantum fluctuations during inflation are included, the relaxion field will have a quantum spread about its classical expectation value. If we require that the quantum spreading  the relaxion undergoes in one e-fold is smaller than the distance it classically rolls down in the same time, we obtain,
\beq
H_I\lesssim \frac{V'(\phi)}{H_I^2}.
\label{qvsc}
\eeq
One can check that the slow roll conditions,  are always satisfied if both \eq{cond1} and \eq{qvsc} are true. Together \eq{deriv}, \eq{cond1} and \eq{qvsc}  imply an upper bound on the cut-off scale,
\begin{eqnarray}
M \lesssim \left(\frac{\Lambda_{br}^4}{f}\sin \frac{\phi_0}{f}\right)^{1\over 6} \sqrt{M_{\rm Pl}}\,
\label{cosmo}
\end{eqnarray}
which gives $M \lesssim 100$ TeV ($M \lesssim 10^9$ GeV) taking $f\gtrsim 10^9$ GeV and $\sin \frac{\phi_0}{f}\sim 10^{-10}$  (taking $\Lambda^4_{br}\sim v^4,~f\gtrsim M$ and $\sin \frac{\phi_0}{f}\sim 1$) for the QCD  (non-QCD)  case.
\begin{figure*}[t!]
\includegraphics[width=0.5\textwidth]{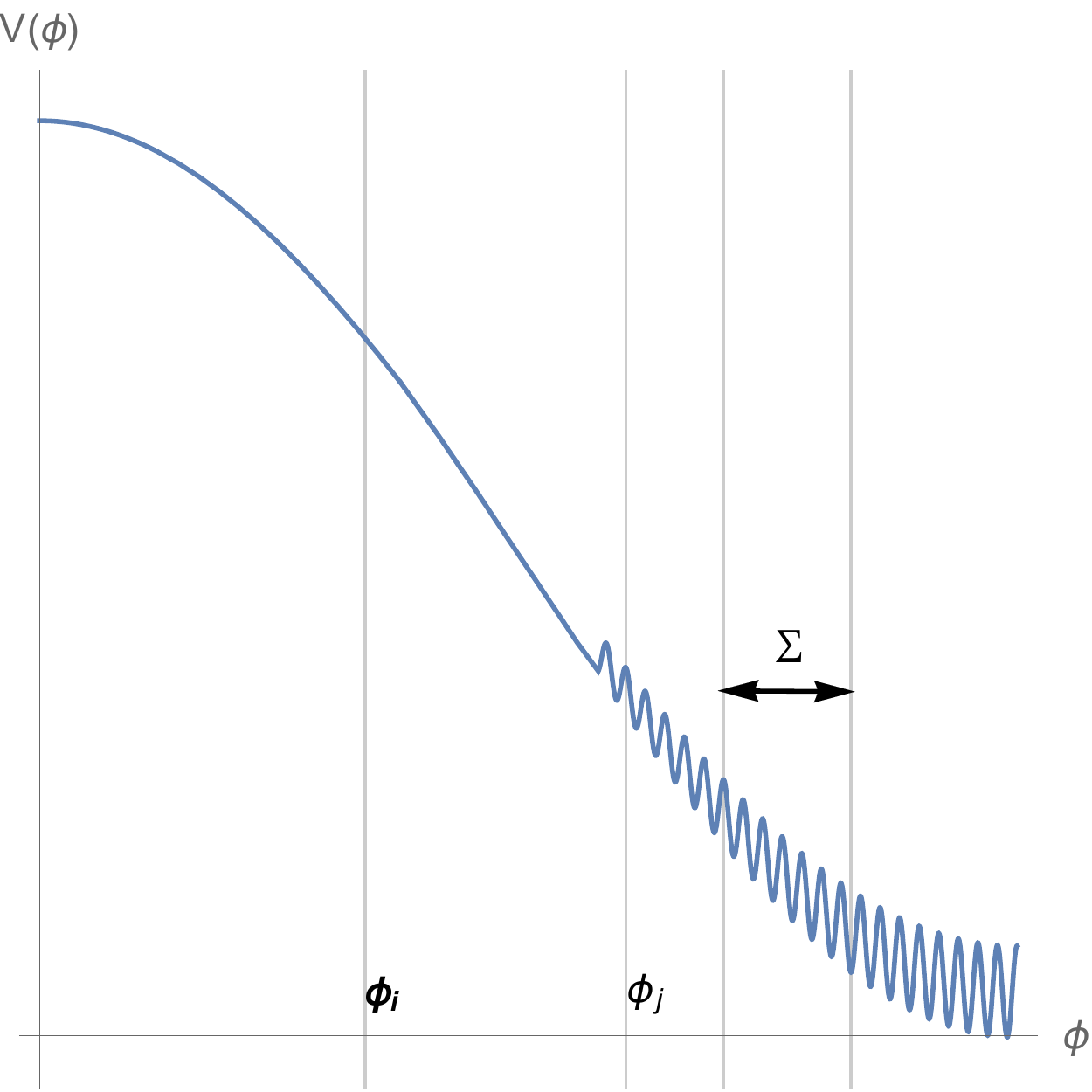}\hfill
\caption{Here is a schematic representation of the relaxion potential in \eq{cosine} after setting $\langle H \rangle^2=-\mu^2(\phi)/\lambda$. The region marked $\Sigma$ shows the vacua where the relaxion  stabilises during inflation taking into account its quantum spreading in \eq{delta}. Even after its expectation value stabilises, there can be regions where the relaxion has undergone a large quantum fluctuation  to a generic point in field space such as $\phi_i$ above. While such fluctuations are rare during inflation, the fact that the vacuum energy at $\phi_i$ is much larger compared to that in the $\Sigma$ region,  can amplify these fluctuations in volume after the inflaton settles down.  In the NP model the size of the backreaction changes after inflation and this figure shows the backreaction potential before this happens. In this model, $\phi_j$ is the position of the minima that have an ${\cal O}(1)$ relaxion stopping phase after the inflaton stabilises and the barriers become large. }
\label{relpot}
\end{figure*}

Coming to the central issue of this paper, it has been argued in Ref.~\cite{Nelson}  that \eq{qvsc}  may be too strict a condition. This is because, as we will soon see in more detail,  the total quantum spread after $N_e$ e-folds is, $\delta \phi\sim\sqrt{N_e} H$.  The typical number of e-folds is given by ${N_e}\sim H_I  \Delta \phi/\dot{\phi} \sim H_I^2/g^2\sim H_I^2F^2 /M^4$ where $\Delta \phi\sim M^2/g\sim F$ is the typical field excursion. Finally  we obtain for the variation in $\phi$ and $\mu$,
\bea
\delta \phi &\sim& \frac{H_I^2}{g}\sim \frac{F H_I^2}{M^2} \nonumber\\
\delta \mu^2 &\sim& H_I^2
\label{delta}
\eea
where to obtain the second line one needs to use  \eq{linear} or \eq{cosine}. Thus at the end of inflation the relaxion field is spread over many vacua but the variation in the electroweak scale  is ${\cal O}(1)$ as long as,
\beq
H_I\lesssim v.
\label{hbound}
\eeq
This region is marked as $\Sigma$ in Fig.~1.

 In the NP model a high Hubble scale is used to evade the problem of an  ${\cal O}(1)$   $\theta_{\rm QCD}$ and thus provide a unified solution to the hierarchy and strong CP problems. The authors use the fact that if the Hubble scale during inflation is larger than a few GeV,  the associated Gibbons-Hawking temperature~\cite{Gibbons} suppresses the usual axion potential of  \eq{br} by a factor of $1/\theta_{\rm QCD}^{\rm ub}\sim 10^{10}$ or more. As a result,  once the Hubble becomes smaller towards the end of inflation, the backreaction wiggles become larger and the relaxion settles down at a point where the strong CP phase is smaller than ${\cal O}(10^{-10})$ (see \eq{deriv}).     A similar solution to generate a small $\theta_{\rm QCD}$,  also involving a change in the relaxion potential after inflation, was proposed already in Ref.~\cite{gkr} but that mechanism is less elegant as it involves somewhat ad hoc couplings of the relaxion to the inflaton \footnote{Another alternative route to a unified relaxion solution to the hierarchy and strong CP problems is by identifying the ${\cal O}(1)$ relaxion stopping phase with the CKM phase and using the Nelson-Barr mechanism to solve the strong CP problem as in Ref.~\cite{nbr,hir}.}.  Another advantage of relaxing the QvsC condition, true for both the NP model as well as more general non-QCD  models, is that the only bound on the cut-off would now be independent of the backreaction scale and  weaker than \eq{cosmo}, 
\bea
M\lesssim \sqrt{H_I M_{pl}}\sim 10^{10} {\rm~GeV} \sqrt{\frac{H_I}{10^2{\rm~GeV}}}
\label{cutoff2}
\eea
where we have used \eq{cond1}. As we discuss in the next section, however,  large quantum fluctuations  become problematic after the `end' of inflation.

\section{The Relaxion Measure problem}


In this section we give a more  careful treatment of the effect of  quantum fluctuations which will lead us to a precise statement of the  `Relaxion Measure Problem'. We will closely follow the presentation of the measure problem of usual eternal inflation  in Ref.~\cite{linde1,linde2,linde3} except that here we would be concerned with  the quantum fluctuations of the relaxion and not the inflaton. Quantum effects are usually incorporated as a stochastic noise term that  provides a Hubble sized  kick, $\pm H_I$, to the field value in every e-fold. With these quantum effects included the time dependance of  $\phi$ is identical to a particle undergoing Brownian motion in a potential gradient. The probability at a given instant, $P_c(\phi,t)$,  for  the field to lie in an interval  $\left[\phi,\phi+ d\phi\right]$ in unit comoving volume, thus obeys the diffusion equation (see for eg. Ref.~\cite{star,gonlin,star2}),
\bea
\frac{\partial P_c}{\partial t}&=&\frac{\partial}{\partial \phi}\Bigg(\frac{H^{3/2}(\phi,\sigma)}{8 \pi^2}\frac{\partial}{\partial \phi}(H^{3/2}(\phi,\sigma)P_c)\nonumber\\&+&\frac{V'(\phi)}{3 H(\phi,\sigma)} P_c\Bigg).
\label{pc}
\eea
The differential equation for the total proper volume, ${\cal V}(\phi,t)$, \footnote{The distribution, ${\cal V}(\phi, t)$, is proportional to the unnormalized probability distribution, $P_p(\phi,t)$, of Ref.~\cite{linde1,linde2,linde3} which denotes the probability  for  the field to lie in an interval  $\left[\phi,\phi+ d\phi\right]$ in unit physical (proper) volume. }
  having $\phi$ in the range $\left[\phi,\phi+ d\phi\right]$ has the same form~\cite{nambu1,nambu2,linde4},
\bea
\frac{\partial {\cal V}}{\partial t}&=&\frac{\partial}{\partial \phi}\Bigg(\frac{H^{3/2}(\phi,\sigma)}{8 \pi^2}\frac{\partial}{\partial \phi}(H^{3/2}(\phi,\sigma){\cal V})\nonumber\\&+&\frac{V'(\phi)}{3 H(\phi,\sigma)} {\cal V}\Bigg)+3 H(\phi,\sigma) {\cal V}
\label{pp}
\eea
 apart from the last term above that takes into account the Hubble expansion of the volume element per unit time. This  last term will play crucial role in this work  as it will allow regions with large quantum fluctuations and a small $P_c(\phi,t)$ to still have a large volume, ${\cal V}(\phi, t)$,  if the  vacuum energy in these regions is large.    Without  this  term ${\cal V}(\phi, t)$ would be proportional to  $P_c(\phi,t)$, as is the case during inflation when  the relaxion has a sub-dominant contribution to the energy density and  the Hubble scale is independent of $\phi$ . We thus find that solutions to  \eq{pc} and \eq{pp}  are related in a simple way,
\beq
{\cal V}(\phi,t)= P_c(\phi,t) \frac{e^{\int_0^t 3 H(\langle \sigma(t)\rangle) dt}}{H_I^3}
\label{vprel}
\eeq
for $t<t_i$, $t_i$ being the time when  the inflaton stabilises. To obtain the above relationship we have assumed that the whole universe emerges from a single Hubble patch of size $1/H_I^3$ at $t=0$ and ignored any quantum fluctuation of the inflaton about its  classical trajectory $\langle{\sigma(t)}\rangle$. For simplicity from here on we will assume a  constant Hubble scale  $H(\langle \sigma(t)\rangle)\approx H_I$ for $t\lesssim t_i$. Note that the relaxion stabilises  at a time, $t_r$, that is   intermediate between 0 and  $t_i$ and both $t_{r,i}\sim N_e/H_I\sim H_I/g^2 \sim H_I F^2/M^4$.  We want to  now solve for $P_c(\phi,t)$ and thus ${\cal V}(\phi,t)$ for $t\lesssim t_i$. We first attempt to find the solution ignoring the backreaction. It  is still very difficult to obtain analytically a solution for the cosine potential in \eq{cosine} but it is straightforward in the case of the linear potential in \eq{linear} (see Ref.~\cite{Nelson}), 
\beq
P_c(\phi_i,t_i)=\sqrt{\frac{2 \pi}{H_I^3 t}}\exp{\left(\frac{-2 \pi^2(\phi_i-\langle \phi (t_i) \rangle)^2}{H_I^3 t}\right)}
\label{prob}
\eeq
where we have taken $P_c(\phi_i,0)$ to be a delta function about a generic  initial field value. The expectation value $\langle \phi \rangle$ obeys the classical EoM for $\phi$.  It might seem that using the linear potential in is a bad approximation as the relaxion travels a distance in field space $\sim F$. Using the full potential would, however,  give us a smaller suppression in $P_c$ for large  $(\phi_i-\langle \phi (t_i) \rangle)$ as  the potential gradient, $V'(\phi)$ would be smaller than the linear case (for any $\phi \neq \pi F/2$). Thus we can obtain an  estimate for  $P_c(\phi_i,t_i)$ for the periodic  potential in \eq{cosine}  that is conservative for our purposes by substituting $g\sim M^2/F$ (see \eq{interchange}) in \eq{prob} above.

 The  solution in \eq{prob} strictly applies only for $t\lesssim t_r$, that is it applies until the time  $\langle \phi \rangle$ reaches the backreaction barriers and eventually stabilises at a vacuum with weak scale Higgs VEV  (in the region marked $\Sigma$ in Fig. 1).  For $t_r\lesssim t \lesssim t_i$ while the form of $P_c$  above is still reliable if $(\phi_i-\langle \phi(t_i) \rangle)$ is large as is the case for any point far from the wiggles such as the point $\phi_i$ in Fig. 1, the effect of the backreaction can become significant  in the region with the wiggles.   Note, however, that if the Hubble scale, $H_I$, is taken to be large (as in the NP model)  and the QvsC condition in \eq{qvsc} is violated, the second term in \eq{pp} becomes irrelevant even in the presence of the wiggles. Therefore even though $\langle \phi \rangle$ stops evolving, the quantum spreading of the relaxion  continues as before controlled by the first term in \eq{pp}.  The bottomline is that the quantum spread of the relaxion field at $t\sim t_i$ is not expected to exceed the square root of the variance of the above gaussian evaluated at $t=t_i$, i.e. $\delta\phi \sim \sqrt{H_I^3 t_i} \sim H_I^2/g \sim F H_I^2/M^2$, the region marked $\Sigma$ in Fig. 1,  and thus the variation  in $\mu^2$ is ${\cal O}(1)$ as already derived in \eq{delta}. It follows from our definition of  the region $\Sigma$  that, 
\bea
\int_{\Sigma'} P_c (\phi_i, t_i) d\phi &=&1-\int_\Sigma P_c (\phi_i, t_i) d\phi_i \ll 1\label{sigma},
\eea
where $\Sigma'$ is the whole region in field space outside $\Sigma$. From here on we will abbreviate $P_{\Sigma,\Sigma'}=\int_{\Sigma,\Sigma'} P_c (\phi_i, t_i) d\phi_i$. A more detailed treatment of how the backreaction affects   $P_c$ can be found in Ref.~\cite{Nelson}.

Let us now analyse what happens for $t>t_r$. Consider the patches where $\phi$ has undergone huge quantum fluctuations  and the relaxion field sits at a point such as  $\phi_i$ in Fig. 1. The key point is that after  the inflaton  stabilises, the relaxion energy density is no longer a sub-dominant component of the total energy density  in such patches.  While the probability of such a  large quantum fluctuation  at $t=t_i$ is exponentially suppressed (see \eq{prob}), the volume of such patches would grow exponentially   driven by the ${\cal O}(M^4)$ relaxion energy density.  As we will soon see if the QvsC condition is not imposed the exponential growth can overcome this exponential suppression. This is the central point of this work: one cannot neglect  patches  with large seemingly unlikely quantum fluctuations, i.e. if  $P_c(\phi,t)$ is small,  as they can grow exponentially in physical volume, i.e.  ${\cal V}(\phi,t)$ can still become large. 

Furthermore, in such an eternally inflating universe, the volume fraction of patches with a given property is ambiguous and depends on regularisation. Let us understand this point more carefully in our context. We want to compare, for $t>t_i$,  the relative size of the four volume where the relaxion has stabilised to a vacuum with a small Higgs VEV (and in the case of the NP model also a small $\theta_{\rm QCD}$)  to the four volume where this is not true; we define the ratio of the latter   to the former as follows, \bea
 \xi=\lim_{t_c \to \infty}\frac{{\cal V'}^4(v\ll M, \theta_{\rm QCD} \ll 1, t_i<t<t_c)}{{\cal V}^4 (v\ll M, \theta_{\rm QCD} \ll 1, t_i<t<t_c)}
 \label{xigen}
 \eea
 where the four volumes are defined as the integral over  the invariant volume element $\sqrt{|\det g|} d^4 x$ in the relevant regions.   As both these volumes are generally infinite,  to define their ratio we have to use a time regulator above, i.e we compute the ratio first restricting ourselves to $t<t_c$ and then take the limit $t_c \to \infty$. There is, however, no unambiguous way to introduce the time regulator above and this leads to the measure problem. This is because  there is no unique way to choose global time slices across casually disconnected regions of  spacetime.   The most natural way to define time slices across the different patches is to take the proper time elapsed, $t$,  along the geodesics starting from the initial Hubble patch at $t=0$. This is called the proper time cut-off measure and was first used in~\cite{linde1}. While the proper time cut off measure seems like the most natural choice, it is known  to lead to many paradoxes in the usual case (unlike here) of eternal inflation driven by quantum fluctuations of the inflaton. One of the most promising alternatives is the scale factor cut off measure~\cite{star2, linde2, linde3} which manages to evade many of these issues by choosing  constant scale-factor time slices~\cite{desimone1, bousso, desimone2}.
 
For the NP model we can rewrite \eq{xi} in an interesting way where the denominator in \eq{xi} corresponds to the the four volume, ${\cal V}_\Sigma^4$,  generated by the expansion of regions where the field value lies in the $\Sigma$ region at $t=t_i$   whereas the numerator corresponds to the  four volume, ${\cal V}_{\Sigma'}^4$,  generated from the region outside $\Sigma$ at $t=t_i$ (see Fig. 1).  This is because, as we will show in the following sections, for the NP model ${\cal V}_{\Sigma'}^4$  contains regions where either the Higgs VEV does not stabilise at all or regions where $\theta_{\rm QCD}\sim 1$. Thus we obtain, 
 \bea
 \xi=\lim_{t_c \to \infty}\frac{{\cal V}^4_{\Sigma'} (t_c)}{{\cal V}^4_{\Sigma}(t_c)}~~~~~{\rm(NP~model)}.
 \label{xi}
 \eea
As we will see later, in general relaxion models even ${\cal V}^4_{\Sigma'}$ can contain regions with weak scale Higgs VEV so that for such cases \eq{xigen} does not imply \eq{xi}.

In the following sections we will compute  $\xi$ in both these measures and for both the NP model as well as more general (non-QCD) relaxion models where the size of the backreaction does not change after the inflaton stabilises.  If  $\xi \lesssim 1$ it will indicate that the relaxion mechanism can overcome the issues raised in this work. Before going into the details of the calculation let us clarify an important issue about the definition of $\xi$ in \eq{xigen}.  Note that in our definition we have not required that the denominator of \eq{xigen}  contain patches with a small cosmological constant (CC) in addition to having a small Higgs VEV (and in the case of the NP model also a small $\theta_{\rm QCD}$). This is because   in relaxion models the CC problem is solved by  tuning, i.e.  $\Lambda_{c}$ in \eq{linear} or \eq{cosine}  is tuned to almost exactly cancel the vacuum energy  in one of the vacua in the $\Sigma$ region, and this happens to be the vacuum we live in. Thus as long as $\xi \lesssim 1$,  it means that most of the physical volume has a weak scale VEV  (and in the case of the NP model also an acceptable $\theta_{\rm QCD}$) and the further  requirement of having the correct CC  is achieved by brute force tuning.

\section{Proper Time Cut-off Measure} 
We now estimate  $\xi$ in the proper time cut-off measure.  Our computation will depend on whether the relaxion dynamics is classical or quantum for  $t\gtrsim t_i$.  This splits the paramerter space into two regions depending on the Hubble scale that now gets a contribution only from the relaxion vacuum energy, 
\bea
H(\phi)=\frac{1}{M_{pl}}\sqrt{\frac{V(\phi)}{3} }.
\label{hubble}
\eea
The two regions are, 
\bea
&&{\rm Case~(i):}~H(\phi_m) \lesssim (V(\phi))^{1/3}\nonumber\\
&&{\rm Case~(ii):}~H(\phi_m) \gtrsim (V(\phi))^{1/3}\label{cases}
\eea
where $\phi_m=0$ is the point with maximal vacuum energy. In the first regime, the dynamics of the relaxion is always classical  and the first term in \eq{pp} can be ignored  whereas in the second regime quantum diffusion effects  encoded in this term cannot be ignored. The condition, $H(\phi_m) \lesssim (V(\phi))^{1/3}$, can be rewritten as the upper bound on the cut off in \eq{cosmo} which applies to Case (i) whereas the cut-off is bounded only by \eq{cutoff2} for Case (ii) and can thus be much higher.

\subsubsection{CASE (i): Classical dynamics  for $t>t_i$} 

\paragraph{NP Model}  Let us first estimate  denominator in \eq{xi}, ${\cal V}^4_{\Sigma}(t_c)$.  To compute this volume we need to know the expansion rate of a typical vacuum in the $\Sigma$ region. Recall that the CC   is tuned to the observed value in one of the vacua in the  $\Sigma$ region  (our vacuum). This implies that regions in the vacua below this particular one in Fig.~1 will have a negative CC and would thus collapse and not contribute to ${\cal V}^4_{\Sigma}(t_c)$ for large $t_c$.  On the other hand the vacuum energy of a typical vacuum with positive CC in this region is $\delta \mu^2 M^2\sim H_I^2 M^2$ (see \eq{delta}) and the corresponding Hubble scale is thus  $H_\Sigma \sim H_I M/M_{pl}$. We thus obtain, 
\beq
{\cal V}^4_{\Sigma}(t_c) \sim \frac{e^{3 N_e} P_\Sigma }{H_I^3} \int_{t_i}^{t_c} e^{3 H_\Sigma t} dt=\frac{P_\Sigma e^{3 N_e}  e^{3 H_\Sigma (t_c-t_i)}}{3 H_I^3 H_\Sigma}.
\label{denom}
\eeq
where the pre-factor before the integral is the three dimensional volume of the $\Sigma$ region at $t=t_i$ (see \eq{vprel}). We have omitted in our estimate above an ${\cal O}(1)$ factor to account for the fact that the fraction of vacua in  $\Sigma$ with negative CC do not contribute to ${\cal V}^4_{\Sigma}(t_c)$.

Now we compute the four volume, ${\cal V}_{\Sigma'}$ in the NP model. The first term in the  right hand side of \eq{pp} is negligible for this case and the relaxion dynamics is classical for $t\gtrsim t_i$. The relaxion will classically slow roll but now with a field dependent Hubble friction given by \eq{hubble}. For the NP model  one can check that  $H(\phi) \ll \Lambda_{\rm QCD}$ for all $\phi$,  so that  the barriers are now large given by \eq{brqcd}. As a result   if the relaxion field starts to roll from the point  $\phi_i$  in a Hubble patch at $t=t_i$,   it stops at a point $\phi_j$  at a later time $t_j(\phi_i)$ (see Fig. 1), where $\theta_{\rm QCD}=\phi_j/f$ is ${\cal O}(1)$ and the Higgs VEV, $\langle H\rangle <\theta_{\rm QCD}^{\rm ub} v \sim 10^{-10} v$ by  \eq{deriv}.  At this point the Hubble scale $H_j=H(\phi_j)\sim v M/M_{pl}$.  One can verify that the slow roll conditions are satisfied in the full range while the field rolls from $\phi_i$ to $\phi_j$.   Once the subdominant quantum corrections are taken into account the relaxion in these patches would stop in a small region around $\langle \phi \rangle \sim \phi_j$. Assuming a homogeneous universe where the relaxion field takes its classical value everywhere,  the total four volume that a single patch at $t=t_i$ with $\phi=\phi_i$ inflates into by the time $t_c \gg t_{i,j}$ is given by,
   \bea
 \delta {\cal V}_{\Sigma'}(\phi_i,t_c)&=&\exp{\left(3\int_{t_i}^{t_j(\phi_i)} H(\langle \phi(t)\rangle) dt\right)} \frac{e^{3 H_j (t_c-t_j(\phi_i))}}{3 H_I^3 H_j}  \nonumber\\
 &=& \exp{\left(9\int_{\phi_j}^{\phi_i} \frac{H^2({\phi})}{V'({\phi})}d{\phi}\right)} \frac{e^{3 H_j (t_c-t_j(\phi_i))}}{3 H_I^3 H_j}
\label{delv1}
   \eea
where the first exponential factor is due to the volume growth  as the field rolls from from $\phi_i$ to $\phi_j$ and the second exponential factor is the volume growth due to the fixed cosmological constant for $t>t_j$.   $\langle {\phi}(t)\rangle$ is the solution to the EoM, $\dot{\phi}=-V'(\phi)/3H(\phi)$, a fact we use to arrive at the second line above. The $1/H_j$ factor arises from the integral of the three dimensional volume over time in the large $t_c$ limit. The argument of the first exponential factor can be  evaluated for the periodic potential in \eq{cosine}, 
\beq
 9\int_{\phi_j}^{\phi_i} \frac{H^2({\phi})}{V'({\phi})}d{\phi}=\frac{3 F^2}{M_{pl}^2}\int_{\phi_i/F}^{\phi_j/F} \frac{\cos x+\alpha}{\sin x} dx=\frac{3 F^2}{M_{pl}^2} I\left(\frac{\phi_i}{F}\right)
 \label{inte}
\eeq
  where $\alpha=\Lambda_c/\kappa_2 M^4$.  As the four volumes arising from the different Hubble patches at $t=t_i$ are casually disconnected, we can find the four volume ${\cal V}^4_{\Sigma'}$  simply by convoluting $\delta {\cal V}^4_{\Sigma'}(\phi_i,t_c)$  with the probability distribution $P_c(\phi_i, t_i)$ in \eq{prob},
  \bea
{\cal V}^4_{\Sigma'}(t_c)&=& e^{3 N_e}\int_{\Sigma'}P_c(\phi_i, t_i)\delta{\cal V}^4_{\Sigma'}(\phi_i,t_c) d\phi_i.
\label{conv}
  \eea
   Mathematically this is true because, being a linear equation,   \eq{pp} can be first solved  with  the initial condition  that ${\cal V}(\phi_i,t_i)$ is a delta function around $\phi_i$ and  the solutions for the different $\phi_i$ can then be superposed. For $P_c(\phi_i, t_i)$ we will use \eq{prob} replacing $g\sim M^2/F$ as explained in the previous section.  Now using Eq.~(\ref{xi},~\ref{denom},~\ref{delv1},~\ref{inte},~\ref{conv}) we obtain for $t_c \gg t_i$ the lower bound,
   \bea
\frac{{\cal V}^4_{\Sigma'}(t_c)}{{\cal V}^4_{\Sigma}(t_c)}&\gtrsim&{\cal K}\int_{\Sigma'}\exp{\left(\frac{3 F^2}{M^2_{pl}}I(x_i)-\left(\frac{\sqrt{2} \pi M^2}{H_I^2}\right)^2  \Delta x_i^2\right)} dx_i \nonumber\\ &\times&\frac{H_\Sigma}{H_j}\frac{e^{3 H_j (t_c-t_j(\phi_m))}}{e^{3 H_\Sigma (t_c-t_i)}}
\label{2fac}
  \eea
  where   ${\cal K}= {\sqrt{2 \pi} M^2}/{(P_\Sigma  H_I^2)}$ and $\Delta x_i=x_i-\langle x_i \rangle$.   The first line on the right hand side above denotes the ratio of the total number of patches that end up in vacua near $\phi_j$  to the number of patches that end up the region $\Sigma$.  The second line is a lower bound on the ratio of the four volume generated from a single patch at $\phi_j$ to that generated from a single patch in the $\Sigma$ region; here we have used the fact that the time it takes for a the field value  to roll down from  the maxima of the potential, $\phi_m=0$, to $\phi_j$ is greater than the corresponding time for any other starting field value, i.e. $t_j(\phi_m)\geq t_j(\phi_i)$. We now show that both these factors are greater than unity in the NP model so that $\xi >1$.  Taking $f\gtrsim 10^9$ GeV  and using \eq{deriv}, one can check that  a transplanckian $F\gg M_{pl}$ is required to achieve a  cut-off above weak scale in the NP model. Thus the coefficient of $I(x_i)$ above is much larger than 1.   Therefore, as both $\Delta x_i^2$ and $I(x_i)$ are ${\cal O}(1)$, the factor in the first line of \eq{2fac} is exponentially large  unless the  coefficient of the  $\Delta x_i^2$  is larger than the coefficient of $I(x_i)$. The latter condition can be rewritten as follows, 
  \beq
 \frac{H_I^2M^2}{M_{pl}}\lesssim \frac{M^4}{F} \sim \frac{y_u v f_\pi^3 \theta_{\rm QCD}}{f}
 \label{newcond}
 \eeq
where we have used \eq{deriv} for the last step.  If the QvsC condition is violated, to the extent it is in the NP model where we must have $H_I>3$ GeV, \eq{newcond} is never satisfied for any cut-off value larger than the weak scale.  Thus the number of patches where the relaxion eventually  stabilises around $\phi_j$ is exponentially larger than the number where it stabilises in the $\Sigma$ region in Fig.~1. Let us now come to the factor in the second line. Note that although $H_\Sigma$ can be of the same order as $H_j$ for the maximal value $H_I\sim v$ (see \eq{hbound}), one always has $H_j>H_\Sigma$  as the region around $\phi_j$ where $\theta_{\rm QCD}\sim 1$ can strictly not overlap with the region $\Sigma$ where where $\theta_{\rm QCD} \ll 1$. Thus clearly the second line in \eq{2fac} is also greater than unity for a large enough $t_c$. 

To summarise, for this case using the proper time cut-off measure we find that  the NP model predicts  that the number of patches where the relaxion stabilises in the vacua near $\phi_j$ (where $\theta_{\rm QCD} \sim 1$) are exponentially larger than the patches where the relaxion stabilises in the $\Sigma$ region.   Subsequently the patches with field value around $\phi_j$ expand at a faster rate because of the larger value of the Hubble scale relative to that in the $\Sigma$ region increasing further the ratio in \eq{2fac}. This gives a divergent  $\xi$ in \eq{xi} so that the relaxion mechanism does not work as intended for this case. 

\paragraph{General relaxion models}  We now consider   the fate of more general relaxion models if there is no restriction on $H_I$ such as the QvsC condition. Consider first the four volume emerging from  the $\Sigma'$ region.  As we are in the classical regime with  $H(\phi_m) \lesssim (V(\phi))^{1/3}$, after the time $t=t_i$ the relaxion field value starts classically rolling down from the $\Sigma'$ region towards the minima in the $\Sigma$ region. Even though the dynamics is well approximated classically, the final fate of the relaxion  depends in an important way on the    quantum spread  of the relaxion,   $\delta \phi \sim M^4/ (g M^2_{Pl}) \sim M^2F/M^2_{Pl}$, that nevertheless exists.  The relaxion can stop at any vacuum  above the vacuum with the tuned CC, as long as the vacuum energy is large enough to ensure that the  slow roll conditions,
\bea\label{sroll}
\varepsilon&=&\frac{M_{Pl}^2}{2} \left( \frac{V'(\phi)}{V(\phi)}\right)^2\ll1 \nonumber\\
\eta&=&M_{pl}^2  \frac{V''(\phi)}{V(\phi)}\ll 1
\eea
are not violated.  It cannot, however,   stop at the minimum with the tuned CC.  This is because as it approaches this minimum the total vacuum energy and thus the Hubble friction  vanishes leading to a violation of  the slow roll conditions above. As a result the relaxion field shoots  past this minimum with non-zero kinetic energy.  Parts of the universe where this occurs keep inflating and the total energy keeps decreasing until it becomes zero. Thereafter these regions   collapse~\cite{fastroll}.  Thus eventually ${\cal V}^4_{\Sigma'}$  contains only the regions where the relaxion stabilises at a vacuum in the $\Sigma$ region with weak scale Higgs VEV  but  a large CC as it stops at a point  necessarily above the vacuum where the CC is tuned; this may be interpreted as a worsening of the CC problem  as the total volume with the correct CC is now  a smaller fraction of the total volume compared to the GKR picture where only ${\cal V}^4_{\Sigma}$ survives.  Therefore, one might  want to limit  the volume, ${\cal V}^4_{\Sigma'}$ (as it necessarily leads to a large CC) and require that   the only volume that survives in the far future is  ${\cal V}^4_{\Sigma}$ as in the original GKR picture.  To derive the condition for this, note that we can get an upper bound on ${\cal V}^4_{\Sigma'}$ by assuming that it contains only regions where the relaxion gets stuck at a vacuum above the one with the tuned CC  and by ignoring the possibility that it can reach the region with negative CC.  This allows us to  recast the computation in  \eq{delv1}-\eq{conv} for this scenario  if we keep in mind that the relaxion stops at a field value $\phi_\Sigma \in \Sigma$ and thus replace $\phi_j \to \phi_\Sigma$. Thus we obtain in the large $t_c$ limit,
 \bea
\frac{{\cal V}^4_{\Sigma'}(t_c)}{{\cal V}^4_{\Sigma}(t_c)}&\lesssim&{\cal K} \int_{\Sigma'}\exp{\left(\frac{3 F^2}{M^2_{pl}}I(x_i)-\left(\frac{\sqrt{2} \pi M^2}{H_I^2}\right)^2  \Delta x_i^2\right)} dx_i \nonumber
\label{2facg}
  \eea  
  where  we must again  replace  $\phi_j \to \phi_\Sigma$ in the definition of $I(x_i)$ and $ \Delta x_i$. For ${\cal V}^4_{\Sigma}$   we have  directly used the result in \eq{denom} which holds here also. The  above ratio is definitely small if,
   \beq
 \frac{H_I^2M^2}{M_{pl}}\lesssim \frac{M^4}{F} \sim \frac{\Lambda_{br}^4}{f}.
 \label{newcondg}
 \eeq
The above inequality is automatically satisfied if we use the condition in \eq{cond1}, $H_I>M^2/M_{Pl}$  in conjunction with the QvsC condition, $H_I<(V'(\phi))^{1/3}$.   \eq{newcondg} implies further that the QvsC condition can be violated  but this gives a new bound on the cut-off $M$. Taking $f\gtrsim M$, a necessary requirement for theoretical consistency~\cite{higgsmix}, and $H_I \lesssim v$ (see \eq{hbound}), we obtain this upper bound,
   \beq
 M \lesssim \left( \frac{M_{pl}\Lambda_{br}^4}{v^2} \right)^{1/3} \sim\left(\frac{\Lambda_{br}}{v}\right)^{4/3}10^8 {\rm~GeV}
 \label{newcutbound}
 \eeq
which is  just an order of magnitude smaller than that from  \eq{cosmo}. This suggests that as long as we satisfy this marginally stronger bound on the cut-off, we can saturate the bound on the Hubble scale in \eq{hbound},  and  have $H_I \sim v$. The possibility of having a larger Hubble scale than that considered by GKR can be interesting from the point of view of  model building of the inflation sector in relaxion models. Note that a higher Hubble scale implies a larger Gibbons-Hawking temperature during inflation a fact that may lead to finite temperature effects that can alter, for instance, the backreaction potential.


\subsubsection{CASE (ii): Quantum dynamics for $t>t_i$}

\paragraph{NP model}
We now examine the second regime in \eq{cases} and estimate $\xi$. In this regime the cut off can be higher, bounded only by the relation in \eq{cutoff2}. Note first of all that the computation of ${\cal V}^4_{\Sigma}$ in the previous subsection is again equally valid in this regime and we can use directly the result in \eq{denom}.

To find  ${\cal V}^4_{\Sigma}$ in this case we have to solve the full differential equation in \eq{pp}.  We solve this equation numerically in the range $\phi_m<\phi<\phi_j$. Give the $\phi \to -\phi$  symmetry of the potential it is natural to take reflecting boundary conditions at $\phi=\phi_m=0$, i.e. we impose that the diffusion current (see Ref.~\cite{linde3}) at the peak vanishes,
\bea
J(\phi_m)=\left.\left(\frac{H^{3/2}}{8 \pi^2}\frac{\partial}{\partial \phi}(H^{3/2}{\cal V})+\frac{V'}{3 H} {\cal V}\right) \right|_{\phi=\phi_m}=0. \label{curr0}
\eea
On the other end  there is a classical minimum  at $\phi=\phi_j$ for $t>t_i$ as explained above \eq{delv1}. One can check that around $\phi_j$, $H(\phi_j)$ is small enough such that quantum fluctuations can be ignored. As the current must vanish for a classically rolling field   at a minimum, we impose  the boundary condition $J(\phi_j)=0$. We find that irrespective of the initial conditions at $t=t_i$,  the solution to \eq{pp} soon reaches a steady state with a uniform rate of expansion that  is nearly the peak value corresponding  to the highest possible energy density, $3 H_m=3 H(\phi_m)$. Thus if we start from a volume of size $1/H_I^3$ with  $\phi=\phi_i$  at $t=t_i$, solving for  ${\cal V}(\phi,t)$ we obtain, 
\beq
{\cal V}(\phi,t| \phi_i)= \frac{\psi(\phi_i)\pi(\phi)}{H_I^3} \exp{\left((3-\delta) H_m) t\right)} .
\label{sst}
 \eeq
for any $t>t_k(\phi_i)$ where ($t_k(\phi_i)-t_i$) is the relaxation time it takes to reach the above steady state.  This is exactly what has been observed for other potentials in the context of inflationary models~\cite{linde1,linde2,linde3}.  The universal functions $\psi(\phi_i)$ and $\pi(\phi)$ in \eq{sst} are positive functions normalised such that their integral over the whole field range is unity; we have kept the original notation for $\psi(\phi_i), \pi(\phi)$ from  Ref.~\cite{linde2,linde3}. Here $\delta\ll 1$ leads to the small difference in the expansion rate from the maximal value  $3 H_m$ and its precise value depends on the parameter $g\sim M/F$.  This result can be understood as follows.  For a Hubble patch with an  initial field value $\phi_i$ at $t_i$, that evolves to the value $\phi$ at the time $t$, the history that gives maximal contribution to ${\cal V}(\phi,t|\phi_i)$ is one where the field first migrates to the highest point $\phi_m$, stays there for the maximal possible amount of time before coming down to $\phi$ at the time $t$.~~\cite{linde1,linde2,linde3}.  In this regime it, therefore,  makes a big difference whether we use the linear potential of \eq{linear} or the potential in \eq{cosine} where the energy is bounded from above. Indeed, with \eq{linear} even quantum fluctuations of the relaxion for $t<t_i$ might become problematic. This is  because, as already pointed out in Ref.~\cite{gkr},  for the the unbounded potential in \eq{linear}, the relaxion might fluctuate to points where its energy density exceeds that of the inflaton even before the inflaton stabilises.

To obtain the four volume a single patch at $t=t_i$   with $\phi=\phi_i$  grows into by a time $t_c\gg t_{i,k}$ we integrate \eq{sst} over time,
\bea
 \delta {\cal V}^4_{\Sigma'}(\phi_i,t_c) &=&\int_{t_i}^{t_c} \int_{\phi_m}^{\phi_j} {\cal V}(\phi,t| \phi_i) d\phi dt\nonumber\\
 &=& \frac{\psi(\phi_i)e^{3\tilde{H}_m (t_c-t^{max}_k)}}{3 H^3_I \tilde{H}_m},
 \label{delv2}
\eea
where $\tilde{H}_m=(1-\delta/3)H_m$. To obtain the  final expression above, we have used the fact that the the contribution to the time integral between $t_i$ and $t^{max}_k=\max(t_k(\phi_i))$ can be ignored in the large $t_c$ limit. We can again find ${\cal V}^4_{\Sigma'}(t_c)$ with \eq{conv} and using \eq{denom} finally obtain,
 \beq
\xi\sim \lim_{t_c \to \infty}\frac{e^{3\tilde{H}_m(t_c-t^{max}_k)}}{e^{3H_\Sigma(t_c-t_i)}} \frac{ H_\Sigma e^{3 N_e} \eta}{\tilde{H}_m}. 
 \eeq
 where,
 \bea
\eta&=&\frac{\int_{\Sigma'} \psi (\phi_i)P_c (\phi_i, t_i) d\phi_i}{\int_\Sigma P_c (\phi_i, t_i) d\phi_i}.
\eea
The first term diverges in the limit $t_c \to \infty$ whereas the other terms are independent of $t_c$. Clearly $\xi \to \infty$ and we conclude, that for the NP model,  the relaxion mechanism does not work in this regime either if we use the proper time cut-off measure.

\paragraph{General relaxion models}
For  more general relaxion models  one can solve \eq{pp} for $t>t_i$ in the range $\phi_m < \phi< \phi_e$. Here  $ \phi_e$ is the point in field space beyond which the rolling relaxion field exits the slow roll regime defined by \eq{sroll}~\cite{linde3} and thus inflation ends.  For the cut-off values in this regime (see \eq{cases}), $\phi_e$ turns out to be a point just above   the minimum with the tuned CC, and hence the region $\phi<\phi_e$ includes almost the whole of the region in $\Sigma$ with positive CC. The boundary conditions at $\phi_m$ are same as in \eq{curr0} whereas the boundary conditions to be applied at $\phi_e$ are discussed in detail in Ref.~\cite{linde3}.  Again the solution reaches a steady state as in \eq{sst} in the region $\phi_m < \phi< \phi_e$.  The ratio $\xi$ in \eq{xigen} now depends on the profile of $\pi(\phi)$, i.e.,
\beq
\xi= \frac{\int_{\Sigma'} \pi(\phi) d\phi}{\int_{\Sigma, \phi<\phi_e}\pi(\phi) d\phi}.
\eeq
 The  function  $\pi(\phi)$ is  sharply peaked close to $\phi_m$~\cite{linde1,linde2,linde3}  and  suppressed in the $\Sigma$ region as can one can anticipate from the qualitative discussion below \eq{sst}.   Thus even here we  obtain $\xi \gg 1$ and conclude that the relaxion mechanism fails to achieve its desired goal in this scenario.

\section{Scale factor cut-off measure: a possible solution ?}

Now we calculate $\xi$ in a different time regularisation the so called scale factor cut-off measure. In this measure we take global time slices of constant scale factor, $\hat{t}=\log a$. While the definition of this time coordinate is subtle in full generality~\cite{desimone1}, for   uniform expansion driven by vacuum energy, as is the case  in the various scenarios we consider here,  we simply have,
\beq
\hat{t}_2=\hat{t}_1+\int_{\hat{t}_1}^{\hat{t}_2}  H(t) dt
\eeq
that is the time elapsed  in these coordinates is just   the number of e-folds elapsed. In particular the  at the proper time instant $t_i$  time coordinate, $\hat{t}_i=N_e$. The invariant four volume element that we need to compute ${\cal V}^4_{\Sigma,\Sigma'}$ in these coordinates, is given by, 
\beq
\sqrt{|\det g|}d^4x= d^3x \frac{d\hat{t}}{H(x,\hat{t})}.
\eeq
 Let us now try to find in these coordinates  the expression for  $\xi$ in the various scenarios  considered in the previous section. For this, as an intermediate step,  we will need to compute the  volume ${\cal V}^4_{\Sigma}(\hat{t}_c)$. It is easy to recast the expression of \eq{denom} in these coordinates, 
\bea
{\cal V}^4_{\Sigma}({\hat{t}_c})=\frac{P_\Sigma  e^{3\hat{t}_c}}{3H_I^3 H_\Sigma}
\label{sigsf}
\eea
to obtain an expression valid in all the different scenarios we will consider. 

\subsubsection{CASE (i): Classical dynamics for $\hat{t}>\hat{t}_i$} 

\paragraph{NP Model} The expression for  ${\cal V}^4_{\Sigma'}(\hat{t}_c)$ in Case (i) simplifies significantly  in scale factor coordinates. As described earlier, the expansion of the universe in this case is completely determined by the classical evolution of the relaxion field which is initially  distributed across various Hubble patches following \eq{prob}. The Hubble scale in a volume emerging from a single patch at $\hat{t}=\hat{t}_i$  is completely determined in terms of the starting field value $\phi=\phi_i$. This implies a unified expression, ${\cal V}^4_{\Sigma'}(\hat{t}_c)$, that includes both stages of time evolution before and after $\hat{t}_j(\phi_i)$ (discussed in the previous section) for the NP model,
\bea
{\cal V}^4_{\Sigma'}(\hat{t}_c)&=&\frac{e^{3 N_e}}{H_I^3}\int_{\Sigma'}\int_{\hat{t}_i}^{\hat{t}_c} P_c(\phi_i, t_i) e^{\int_{\hat{t}_i}^{\hat{t}_c} 3 d\hat{t}}  \frac{d\hat{t}}{H(\phi_i,\hat{t})}d\phi_i \nonumber\\
&=& \frac{P_{\Sigma'} e^{3\hat{t}_c}}{3H_I^3 H_j}
\label{classsf}
 \eea
where we have taken $\hat{t}_c$ to be larger than all other timescales and used  the fact that $H(\phi_i,\hat{t}) \to H_j$ as $\hat{t} \to \infty$ to obtain the final expression. Finally using \eq{sigma}, \eq{sigsf} and \eq{classsf}, we obtain,
\beq
\xi\sim\frac{P_{\Sigma'}H_\Sigma}{P_{\Sigma}H_{j} } \ll 1
\eeq
which implies a successful unified relaxion explanation of the strong CP and hierarchy problems as intended in the NP model if we use the scale factor cut-off measure. 

\paragraph{General Relaxion models}   In the more general relaxion models  while the time evolution after $\hat{t}_i$ can still be described classically in this regime,  for regions that are part of ${\cal V}^4_{\Sigma'}$,  the the final fate of the rolling relaxion field depends on its  quantum spreading as described earlier. There are two possibilities: in some regions the relaxion might cross the vacuum where the CC is tuned causing the collapse of these regions while in others it might stabilise at a minimum in the $\Sigma$ region with  a weak scale Higgs VEV but a CC  larger than the observed one. Ignoring any possibility of collapse and assuming that the second possibility is what always happens will thus give us an upper bound,
\bea
{\cal V}^4_{\Sigma'}(\hat{t}_c)&\lesssim& \frac{P_{\Sigma'} e^{3\hat{t}_c}}{3H_I^3 H_\Sigma}.
 \label{sigmagen}
 \eea
Using \eq{sigma}, \eq{sigsf} and \eq{sigmagen} we obtain, 
\beq
{\cal V}^4_{\Sigma'}(\hat{t}_c) \ll {\cal V}^4_{\Sigma}(\hat{t}_c),
\label{lthan}
\eeq
where we have used the first line of \eq{classsf} which applies to this scenario as well. Although  \eq{xi}  does not apply to this case, it is still true that  $\xi \ll 1$ given \eq{lthan}; this is because while ${\cal V}^4_{\Sigma'}$ contributes to both the numerator and denominator  in \eq{xigen}, ${\cal V}^4_{\Sigma}$ only contributes to the numerator. Another way of stating this is that in the far future the universe is dominated by ${\cal V}^4_{\Sigma}$ as intended in the original construction of GKR and the problems discussed in this work do not arise.

\subsubsection{CASE (ii): Quantum dynamics for $\hat{t}>\hat{t}_i$} 

\paragraph{NP Model} We now turn to the quantum diffusion regime in \eq{cases}.  For the NP model, we can again solve \eq{pp}  in this time parametrisation (see Ref.~\cite{linde3}) with  the same boundary conditions as before at $\phi=\phi_{m,j}$, i.e. requiring that the probability current vanishes at these two points.  For an initial volume $1/H_I^3$ at $\hat{t}=\hat{t}_i$ with field value $\phi=\phi_i$, the solution   again approaches  a steady state after a time $\hat{t}_{\hat{k}}$, 
\beq
{\cal V}(\phi,\hat{t}\geq\hat{t}_{\hat{k}}| \phi_i)= \frac{\hat{\psi}(\phi_i)\pi(\hat{\phi})}{H_I^3} \exp{(\left(3-\kappa) \hat{t}\right)}.
\label{sst2}
 \eeq
 Here $\kappa \ll 1$ and again its value depends on $g \sim M^2/F$. Again $\hat{\psi}(\phi_i)$ and $\hat{\pi}(\phi)$ in \eq{sst} are positive functions normalised such that their integral over the whole field range is unity;  this implies in particular that $\hat{\psi}(\phi_i)<1$ for all $\phi_i$, a fact we will soon require. We obtain the four volume a single patch at $\hat{t}=\hat{t}_i$   with $\phi=\phi_i$  grows into by a time $\hat{t}_c\gg \hat{t}_{i,\hat{k}}$ by integrating \eq{sst2} over time,
\bea
 \delta {\cal V}^4_{\Sigma'}(\phi_i,\hat{t}_c) &=&\int_{\hat{t}_i}^{\hat{t}_c} \int_{\phi_j}^{\phi_m} {\cal V}(\phi,\hat{t}| \phi_i) d\phi \frac {d\hat{t}}{H(\phi)}\nonumber\\
 &\lesssim& \frac{\hat{\psi}(\phi_i)e^{3 (\hat{t}_c-\hat{t}^{max}_{\hat{k}})}}{3 H^3_I H_\phi}
 \label{delv2sf}
\eea
where $\hat{t}^{max}_{\hat{k}}=\max(\hat{t}_{\hat{k}}(\phi))$,  $H_\phi^{-1}= \int_{\phi_i}^{\phi_m} \frac{\hat{\pi}(\phi)}{H(\phi)} d \phi \sim M_{Pl}/M^2$, and we get the upper bound in the second line because we ignore $\kappa$. For obtaining the final result  we have ignored the contribution to the integral in the finite interval between $\hat{t}_{i}$ and $\hat{t}_{\hat{k}}$ that is negligible given $\hat{t}_c\gg \hat{t}_{i,l}$. We can again convolute with the initial probability distribution, $P_c(\phi_i, \hat{t}_i)=P_c(\phi_i, t_i)$ as in \eq{conv} to obtain, 
\bea
{\cal V}^4_{\Sigma'}(\hat{t}_c)&=& \frac{e^{3\hat{t}_c}e^{-3 (\hat{t}^{max}_{\hat{k}}-\hat{t}_i)}}{3 H_I^3 H_\phi}\int_{\Sigma'} \hat{\psi} (\phi_i)P_c(\phi_i, \hat{t}_i) d\phi_i \nonumber\\
\label{sigp2}
 \eea
 which gives using $H_\phi>H_\Sigma$, $\hat{t}^{max}_{\hat{k}}>\hat{t}_i$, $\hat{\psi}(\phi_i)<1$, \eq{sigma} and \eq{sigsf}, 
 \beq
\xi<\frac{H_\Sigma}{H_{\phi}} \frac{\int_{\Sigma'} \hat{\psi} (\phi_i)P_c (\phi_i, t_i) d\phi_i}{\int_{\Sigma} P_c (\phi_i, t_i) d\phi_i} \ll 1.
\eeq
so that again the relaxion mechanism works successfully for this case. To obtain \eq{sigp2} we have used $\hat{t}_i=N_e$.

\paragraph{General Relaxion models} For general relaxion models our strategy will again  be the same as described below \eq{lthan}: we will show ${\cal V}^4_{\Sigma'} \ll {\cal V}^4_{\Sigma}$ in the far future and thus $\xi \ll 1$.  In order to compute ${\cal V}^4_{\Sigma'}$, we solve \eq{pp}  in the scale factor parametrisation (see Ref.~\cite{linde3}) in the range $\phi_m<\phi<\phi_e$, as we did for the proper time cut-off measure, however at $\hat{t}=\hat{t}_i$ we turn on the probability distribution $P_c(\phi_i, \hat{t}_i$) only in the $\Sigma'$ region. The boundary conditions at $\phi=\phi_e$ have again been discussed in Ref.~\cite{linde3}. Taking $\phi_j\to \phi_e$, the expressions in \eq{sst2} and \eq{delv2sf} are also valid for this case albeit with a different $\hat{\psi},\hat{\pi}$, $\kappa$ and $\hat{t}^{max}_{\hat{k}}$. We obtain again using  $H_\phi>H_\Sigma$, $\hat{\psi}(\phi_i)<1$,  $\hat{t}^{max}_{\hat{k}}>\hat{t}_i$,  \eq{sigma} and \eq{sigsf},
\bea
{\cal V}^4_{\Sigma'}(\hat{t}_c)&=& \frac{e^{3\hat{t}_c}e^{-3 (\hat{t}^{max}_{\hat{k}}-\hat{t}_i)}}{3 H_I^3 H_\phi}\int_{\Sigma'} \hat{\psi} (\phi_i)P_c (\phi_i, t_i) d\phi_i\nonumber\\&\ll& {\cal V}^4_{\Sigma}(\hat{t}_c) 
\eea
for large $\hat{t}_c$. Again only the original GKR region  ${\cal V}^4_{\Sigma}(\hat{t}_c)$  survives in the far future, $\xi \ll 1$  and the relaxion mechanism can be successful.  This is especially interesting  because in this regime the allowed cut-off (see \eq{cutoff2}) can be higher and independent of the backreaction.
 
%
%
%
%
%
%
%

\section{Conclusions}

In this work we explored the implications of not imposing the QvsC condition on  relaxion dynamics and thus allowing large quantum fluctuations. The specific issue we investigate can be understood by  considering a patch where the relaxion field has undergone a large quantum fluctuation and has a field value, such as $\phi_i$ in Fig.~1, far from its classical expectation value in the $\Sigma$ region. While the probability of such large quantum fluctuations are exponentially suppressed, the vacuum energy of the relaxion field in such patches is larger than that in `typical' patches (in the $\Sigma$ region). After the inflaton stabilises this vaccum energy is no longer subdominant and can  lead to an exponentially large expansion rate in such `atypical' patches which can eventually compensate for the initial exponential suppression. The regions arising from expansion of such patches may not have desirable properties such as a small Higgs VEV (or in the NP model a small $\theta_{\rm QCD}$) and thus this feature can potentially derail the relaxion mechanism. Furthermore, the relative size of this dangerous volume depends crucially on the way we regulate the time coordinate and this leads to the `Relaxion Measure Problem'.

We first investigate this issue in the   proper time cut-off measure for both general relaxion models and the NP model of Ref.~\cite{Nelson}. In general models, we find that the original QvsC condition imposed by GKR is sufficient to make this dangerous volume negligible. A positive outcome of our study is that it suggests that the QvsC can be violated without compromising the success of the relaxion mechanism if we accept a marginally stronger bound on the cut-off. This can potentially allow a Hubble scale as large as the weak scale in general relaxion models which can be important for model building of the inflation sector of these models. In the NP model, where a high Hubble scale is required, we find that the relaxion mechanism does not seem to work if one adopts the proper time cut-off measure. This fact was already anticipated by NP and thus they proposed   using the scale factor cut-off measure to resolve this issue. We confirm this suggestion and find that indeed the issue raised here do not arise either for the NP model or in more general relaxion models where the QvsC condition is violated. This has the important implication that both for the NP model and in the more general case, the allowed  cut-off can be independent of the backreaction scale and higher (around $10^{10}$ GeV) than the usual bound derived assuming the QvsC condition. While it is very interesting  that the problems highlighted in this work are rendered harmless in the scale factor cut-off measure, one should keep in mind that this does not resolve the issues completely as at present there is no a priori reason to choose  one measure over another.

\section*{Acknowledgements}
We would like to thank  A.~Nelson, C.~Prescod-Weinstein, J.~L.~Evans and L.~Ubaldi for helpful discussions.

  \end{document}